\documentclass [journal]{IEEEtran}
\usepackage {graphicx}
\usepackage{amsmath}
\usepackage{subfigure}

\begin{document}

% Nombres y direcciones de los autores

\title{A Unifying Statistical Model for Atmospheric Optical Scintillation}

\author{
    \authorblockN{Antonio Jurado-Navas, Jos\'e Mar\'ia Garrido-Balsells, Jos\'e Francisco Paris and Antonio Puerta-Notario\\}
    %\authorblockA{Dpt. Communications Engineering, University of M\'alaga\\
    %Campus de Teatinos s/n, E-29071 M\'alaga, Spain\\
    %Email: \{navas,jmgb, paris,apn\}@ic.uma.es}}
    \thanks{This work was supported by the Spanish Ministerio de Educaci\'on y Ciencia, Project TEC2008-06598.}
    % <-this % stops a space
    \thanks{The authors are with
    the Department of Communications Engineering, University of M\'alaga\, %Campus de Teatinos s/n,
    %E-29071 M\'alaga,
    Spain,
    (e-mail: \{navas,jmgb,paris,apn\}@ic.uma.es).}
    %\thanks{ EDICS: CL1.6 (Optical Communications)}
}

\markboth{}{Jurado-Navas et all: A Unifying Statistical Model for Atmospheric \ldots}

\maketitle\bstctlcite{agz:BSTcontrol}

\begin{abstract}
In this paper we develop a new statistical model for the irradiance fluctuations of an unbounded optical wavefront (plane and
spherical waves) propagating through a turbulent medium under all irradiance fluctuation conditions in homogeneous, isotropic
turbulence. %The proposed model is based on the following paradigms: first, a field following a shadowed Rice model constituted by
%a line-of-sight (LOS) component and a double-component for the scattering: one of them connected and the other one independent to
%the LOS component; such form of field is modulated by a multiplicative factor that obeys lognormal statistics, but approximated
%by a gamma distribution for mathematical convenience, being this modulation process the second paradigm of the model.
The major
advantage of the model is that leads to closed-form and mathematically-tractable expressions for the fundamental channel
statistics of an unbounded optical wavefront under all turbulent regimes. Furthermore, it unifies most of the proposed
statistical models for the irradiance fluctuations derived in the bibliography providing, in addition, an excellent agreement
with the experimental data.
\end{abstract}

\section{Introduction}
Atmospheric optical communication is receiving considerable attention recently for use in high data rate wireless links.
%\cite{Jua06}-\cite{Zhu02}.
However, even in clear sky conditions, wireless optical links experience fluctuations in both the intensity and the phase of an
optical wave propagating through this medium \cite{And98} due to time varying inhomogeneities in the refractive index of the
atmosphere.
%Such fluctuations can produce an increase in the link error probability limiting the performance
%of communication systems. In this particular scenario, the turbulence-induced fading is called scintillation.
The reliability of an optical system operating in such environment can be deduced from a mathematical model for the probability
density function (pdf) of the randomly fading irradiance signal (scintillation). For that reason, one of the goals in studying
optical wave propagation through turbulence is the identification of a tractable pdf of the irradiance under all intensity
fluctuation regimes.

Over the years, many irradiance pdf models have been proposed with different degrees of success. Perhaps the most successful
models are the log-normal \cite{Zhu02}, the lognormal-Rician (or Beckmann model) \cite{Chu87} and the gamma-gamma \cite{AlH01}.
The scope of the lognormal model is restricted under weak irradiance fluctuations. The Beckmann model, although shows excellent
agreement with
experimental data, %and simulation data concerning the pdf,
however a closed-form solution
for its integral is unknown in addition to its inherent poor convergence properties. Finally, %Andrews et al. introduced the modified Rytov
%theory and proposed
the gamma-gamma pdf was suggested as a reasonable alternative to Beckmann's pdf because of its much more tractable mathematical
model.

Now, through this paper, we propose a new and generic propagation model and, from it, we obtain a new and unifying statistical
model for the irradiance fluctuations. The proposed model is valid under all range of turbulence conditions (weak to strong) and
it is found to provide an excellent fit to experimental data. Furthermore, the statistical model presented here can be written in
a closed-form expression and it contains most of the statistical models for the irradiance fluctuations that have been proposed
in the bibliography. Thus, the utility of our model can be extended to many other areas of application as the study of fading in
radiofrequency channels.

\section{The proposed model of  propagation}% including a new scattering component coupled to the line-of-sight contribution}
\begin{figure}[t]
    \centering
            \includegraphics[width=8cm]{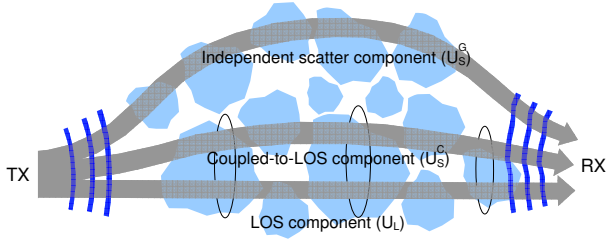}
    \caption{Proposed propagation geometry for a laser beam.% where the observed field at the receiver
%consists of three terms: first, the line-of-sight (LOS) component, $U_L$; the second term is the coupled-to-LOS scattering term,
%$U_S^C$, whereas the third path represents the energy which is scattered to the receiver by off-axis eddies, $U_S^G$.
    }
    \label{FigLosScat}
    \end{figure}

Assume an electromagnetic wave is propagating through a turbulent atmosphere with a random refractive index. As the wave passes
through this medium, part of the energy is scattered and the form of the irradiance probability distribution is determined by the
type of scattering involved. In the physical model we present in this paper, the observed field at the receiver is supposed to
consist of three terms: the first one is the line-of-sight (LOS) contribution, $U_L$, the second one is the component which is
quasi-forward scattered by the eddies on the propagation axis, $U_S^C$, and coupled to the LOS contribution; whereas the third
term, $U_S^G$, is due to energy which is scattered to the receiver by off-axis eddies, this latter contribution being
statistically
independent from the previous two other terms. The inclusion of this coupled to the LOS scattering component %, $U_S^C$,
is the main novelty of the model and it can be justified by the high directivity and the narrow beamwidths of laser beams in
atmospheric optical communications. The model description is depicted in Fig. \ref{FigLosScat}.

Accordingly, we can write the total observed field as:
\begin{equation}
        \begin{split}
            U=&\Bigl(U_L+U_S^C+U_S^G\Bigr)\exp{(\chi+jS)}%=\\
%            =&\Bigl({G}\sqrt{\Omega}\exp{(j\phi_A)}+\sqrt{\rho}{G}\sqrt{2b_0}\exp{(j\phi_B)}+\sqrt{(1-\rho)}U'_S\Bigr)
 %           \exp{(\chi+jS)},
             \label{EqCampoDistribucionM}
        \end{split}
    \end{equation}
\noindent \mbox{where $U_L=\sqrt{G}\sqrt{\Omega}\exp{(j\phi_A)}$, $U_S^C=\sqrt{G}\sqrt{\rho2b_0}\exp{(j\phi_B)}$} and
$U_S^G$=$\sqrt{(1-\rho)}U'_S$, being $U_S^C$ and $U_S^G$ statistically independent stationary random processes. Of course, $U_L$
and $U_S^G$ are also independent random processes. In \eqref{EqCampoDistribucionM}, $G$ is a real variable following a gamma
distribution with $E[G]$=$1$. It represents the slow fluctuation of the LOS component. Following %the notation of
\cite{Abd03}, the parameter $\Omega$=$E[|U_L|^2]$ represents the average power of the LOS term whereas the average power of the
total scatter components is denoted by $2b_0$ = $E[|U_S^C|^2$+$|U_S^G|^2]$. $\phi_A$ and $\phi_B$ are the deterministic phases of
the LOS and the coupled-to-LOS scatter terms, respectively. On another note, $0$$\leq$$\rho$$\leq$$1$ is the factor expressing
the amount of scattering power coupled to the LOS component. Finally, $U'_S$ is a circular Gaussian complex random variable, and
$\chi$ and $S$ are real random variables representing the log-amplitude and phase perturbation of the field induced by the
atmospheric turbulence, respectively. A plausible justification for the coupled-to-LOS scattering \mbox{component, $U_S^C$, is
given in \cite{Ken70}.}

%As an advance, the proposed model offers a highly positive mathematical conditioning due to its obtained irradiance pdf can be
%expressed in a closed-form expression and it approaches as much as desired to the result derived from the lognormal-Rician model,
%for which a closed-form solution for its integral is still unknown. Moreover, it has a high level of generality due to it
%includes as special cases most of the distribution models proposed in the bibliography until now. A plausible justification for
%the coupled-to-LOS scattering component, $U_S^C$, is provided in \cite{Ken70}.

\section{M\'alaga (${\rm{ }}{\cal M}$) probability density function}
From Eq. \eqref{EqCampoDistribucionM}, the observed irradiance of the proposed propagation model can be expressed as:
\vspace{-.2cm}\begin{equation}
    \begin{array}{l}
         {\rm{\emph{I}}} = \Bigl|U_L+U_S^C+U_S^G\Bigr|^2\exp{(2\chi)}=YX, \\
         \hspace{-.15cm}\left\{
            \begin{array}{l}
                 Y \buildrel \Delta \over = \Bigl|U_L+U_S^C+U_S^G\Bigr|^2 \hspace{0.05cm}  ({\hbox{small-scale fluctuations}}) \\
                 X \buildrel \Delta \over = \exp{(2\chi) }  \hspace{1.45cm} ({\hbox{large-scale fluctuations}}), \\
             \end{array}
         \right. %\\
     \end{array}
     \label{EqComponentesIrradiancia}
\end{equation}
\noindent where the small-scale fluctuations denotes the contributions to scintillation associated with turbulent cells smaller
than either the first Fresnel zone or the transverse spatial coherence radius, whichever is smallest. In contrast, large-scale
fluctuations of the irradiance are generated by turbulent cells larger than that of either the Fresnel zone or the so-called
``scattering disk'', whichever is largest.
 From Eq. $\eqref{EqCampoDistribucionM}$, we can rewrite the lowpass-equivalent complex envelope as:
\begin{equation}
        \begin{split}
            \hspace{-1cm}R(t)=&\Bigl(U_L+U_S^C+U_S^G\Bigr)=
            \sqrt{G}\Bigl[\sqrt{\Omega}\exp{(j\phi_A)} \\
            &\hspace{1cm}+\sqrt{\rho}\sqrt{2b_0}\exp{(j\phi_B)}\Bigr]+\sqrt{(1-\rho)}U'_S,
             \label{EqCampoDistribucionM2}
        \end{split}
    \end{equation}
\noindent so that we have the identical shadowed Rice single model employed in \cite{Abd03}, composed by the sum of a Rayleigh
random phasor (the independent scatter component, $U'_S$) and a Nakagami distribution, $\sqrt{G}$. Then, we can apply the same
procedure exposed in \cite{Abd03} consisting in calculating the expectation of the Rayleigh component with respect to the
Nakagami distribution and then deriving the pdf of the instantaneous power. Hence, the pdf of
$Y$ %\buildrel \Delta \over = \Bigl|U_L+U_S^C+U_S^G\Bigr|^2$
is given by:
\begin{equation}
        \begin{split}
            f_Y \left( y \right) = \frac{1}{{\gamma }}\left[ {\frac{{\gamma\beta}}{{\gamma\beta + \Omega^{'} }}} \right]^{\beta} \hspace{-.1cm}
            \exp \left[ { - \frac{y}{{\gamma }}} \right]
            \hspace{-.1cm}~_1F_1\left(\hspace{-.05cm} {\beta;1;\frac{1}{{\gamma}}\frac{\Omega^{'} }{{\left({\gamma\beta  +
            \Omega^{'}} \right)}}y} \right)
             \label{EqPDFY2}
        \end{split}
    \end{equation}
\noindent where
 \mbox{$\beta\buildrel \Delta \over =(E[G])^2/$Var$[G]$} is the amount of fading parameter
 with Var[$\cdot$] as the variance operator, being $| \cdot |$ the absolute value. We have denoted
 \mbox{$\gamma=2b_0(1-\rho)$} and $\Omega^{'}=\Omega+\rho2b_0+2\sqrt{2b_0\Omega\rho}\cos{(\phi_A-\phi_B)}$.
 Finally, $~_1F_1\left(a;c;x\right)$ is the Kummer confluent \mbox{hypergeometric function of the first kind.}

Otherwise, the large-scale fluctuations, $X \buildrel \Delta \over = \exp{(2\chi) }$, is widely accepted to be a lognormal
amplitude \cite{Chu87} but, as in \cite{AlH01, Abd03}, this distribution is approximated by a gamma one, this latter with a more
\mbox{favorable analytical structure. Then:}
    \begin{equation}
        \begin{split}
            f_X \left( x \right) = \frac{{\alpha ^\alpha  }}{{\Gamma \left( \alpha  \right)}}x^{\alpha  - 1} \exp \left(
            { - \alpha x} \right),
        \label{EqPDFX}
        \end{split}
    \end{equation}
\noindent where $\alpha$ is a positive parameter related to the effective number of large-scale cells of the scattering process,
as in \cite{AlH01}.

Now, the statistical characterization of $I$ can be obtained. It is said that $I$ follows a ${\rm{ }}{\cal M}$
distribution %, ${\rm{ }}{\cal M}(\alpha,\beta,\gamma,\rho,\Omega^{'})$,
if $X$ and $Y$ are random variable distributions according to Eqs. $\eqref{EqPDFX}$ and $\eqref{EqPDFY2}$, respectively, with
$\beta$ being a natural number (a generalized expression for $\beta$ being a real number can be also derived, with an infinite
summation, but it is less interesting due to the high degree of freedom of the proposed distribution). Then, the pdf is
represented by:
\begin{equation}
   % \hspace{0.35cm}
        \begin{split}
            f_I \left( I \right) =  A\sum\limits_{k = 1}^\beta  { a_k I^{\frac{{\alpha  + k}}{2} - 1}
            K_{\alpha  - k} \left( {2\sqrt {\frac{{\alpha \beta I}}{{\gamma \beta  + \Omega'}}}
            }~ \right)}
        \label{EqPDFZEspecial}
        \end{split}
    \end{equation}
\noindent where
\begin{equation}
  %  \hspace{1.2cm}
        \left\{
        \begin{split}
        &  A \buildrel \Delta \over = \frac{{2\alpha ^{\frac{\alpha }{2}} }}
                 {{\gamma ^{1 + {\alpha \over 2}} \Gamma \left( \alpha  \right)}}\left( {\frac{{\gamma \beta }}{{\gamma \beta  + \Omega^{'}}}}
                 \right)^{\beta+\frac{\alpha }{2}};  \\
        % ~\\
        &  a_k  \buildrel \Delta \over = \left(\hspace{-.15cm} {\begin{array}{*{20}c}
                                                                {\beta  - 1}  \\
                                                                 {k - 1}  \\
                                                    \end{array}}\hspace{-.15cm} \right)
        {\left( {\gamma \beta  + \Omega^{'}} \right)^{1 - \frac{k}{2}}\over \left( k - 1 \right) !}
        \left( {\Omega^{'}\over \gamma } \right)^{k - 1} \left( {\frac{\alpha }{\beta }} \right)^{\frac{k}{2}}
        .
        \end{split}
         \right.
         \label{EqParametrosPDFZEspecial}
\end{equation}

In Eq. \eqref{EqPDFZEspecial}, $K_{\nu}(\cdot)$ is the modified Bessel function of the second kind and order $\nu$. In the
interest of clarity, the algebraic manipulation to prove this result is moved to Appendix A.

Finally, the $k^{th}$ centered moments of the irradiance signal following a ${\rm{ }}{\cal M}$ distribution, denoted by $m_k
\left( I \right)$, are given by:
\begin{equation}
        \begin{split}
                 m_k \left( I \right) \buildrel \Delta \over =
                E\left[ {I^k } \right] = \frac{{\Gamma \left( {\alpha  + k} \right)}}{{\Gamma \left( \alpha  \right)\alpha ^k }}
                 \frac{1}{\gamma }\left( {\frac{{\gamma \beta }}{{\gamma \beta  + \Omega^{'}}}} \right)^\beta ~ \\
                 \times \sum\limits_{r = 0}^{\beta  - 1} {\beta-1 \choose r}
                 \frac{1}{{r!}}\left( {\frac{{ \Omega^{'}}}{{\gamma \left( {\gamma \beta  + \Omega^{'}} \right)}}} \right)^r
                 \frac{{\Gamma \left( {k + r + 1} \right)}}
                 {{\left( {\frac{\beta }{{\gamma \beta  + \Omega^{'}}}} \right)^{k + r + 1} }}.
             \label{EqMomentsZEspecial}
        \end{split}
    \end{equation}

%For the sake of clarity of the whole paper,
The proof of this result is, again, moved to Appendix B.

To conclude this section, we attach in Table \ref{TableResumen} a list of most of the existing distribution models for
atmospheric optical communications and how they can be generated from the ${\rm{ }}{\cal M}$ distribution model proposed here.
The analytical expression for the different distribution models include in Table \ref{TableResumen} can be consulted in the
references provided next to their names.
\begin{table}[htb]
%    \centering\caption{List of existing statistical models for atmospheric optical communications and generation by using the proposed
 %   ${\rm{ }}{\cal M}$ distribution model.}
     \centering\caption{Generation of different statistical models from the proposed
    ${\rm{ }}{\cal M}$ distribution model.}
    \begin{tabular}{c | cp{3.5cm}}
        \hline
        {Distribution model} & {Generation}  \\ \hline
        {Rice-Nakagami} \cite{And98} & {$\rho=0$} \\  ~ & {Var$[|U_L|]=0$} \\ \hline
        ~ & {$\rho=0$}  \\  {Lognormal} \cite{And98,Str78} & {Var$[|U_L|]=0$} \\ ~ &
                                                            {$\gamma\rightarrow 0$} \\ \hline
        {Gamma} \cite{AlH01} & {$\rho=0$} \\ ~ & {$\gamma=0$} \\ \hline
        {Shadowed-Rician distribution} \cite{Abd03} & {Var$[|X|]=0$} \\ \hline
        {K distribution}\cite{AlH01,Jak80} & {$\Omega=0$ and $\rho=0$} \\ ~ & {or $\beta=1$} \\ \hline
        ~ & {$\Omega=0$} \\ {HK (Homodyned K) distribution} \cite{Jak80} & {$\rho=0$} \\
                                                                      ~ & {$X=\gamma$} \\ \hline
        ~ & {$\Omega=0$} \\ {Exponential distribution} \cite{AlH01} & {$\rho=0$} \\
                                                                    ~ & {$\alpha\rightarrow \infty$} \\ \hline
        {Gamma-gamma} \cite{AlH01}  & {$\rho=1$, then $\gamma=0$} \\ {distribution}
                                                                      & {$\Omega^{'}=1$} \\ \hline
        {Gamma-Rician distribution} \cite{Chu87} & {$\beta\rightarrow\infty$} \\ \hline
    \end{tabular}
    \label{TableResumen}
\end{table}

\section{Comparison with experimental %plane wave and spherical wave
data} 
\begin{figure}[tb]
        \centering
        \subfigure[]{
        \includegraphics[width=4.12cm,height=3.2cm]{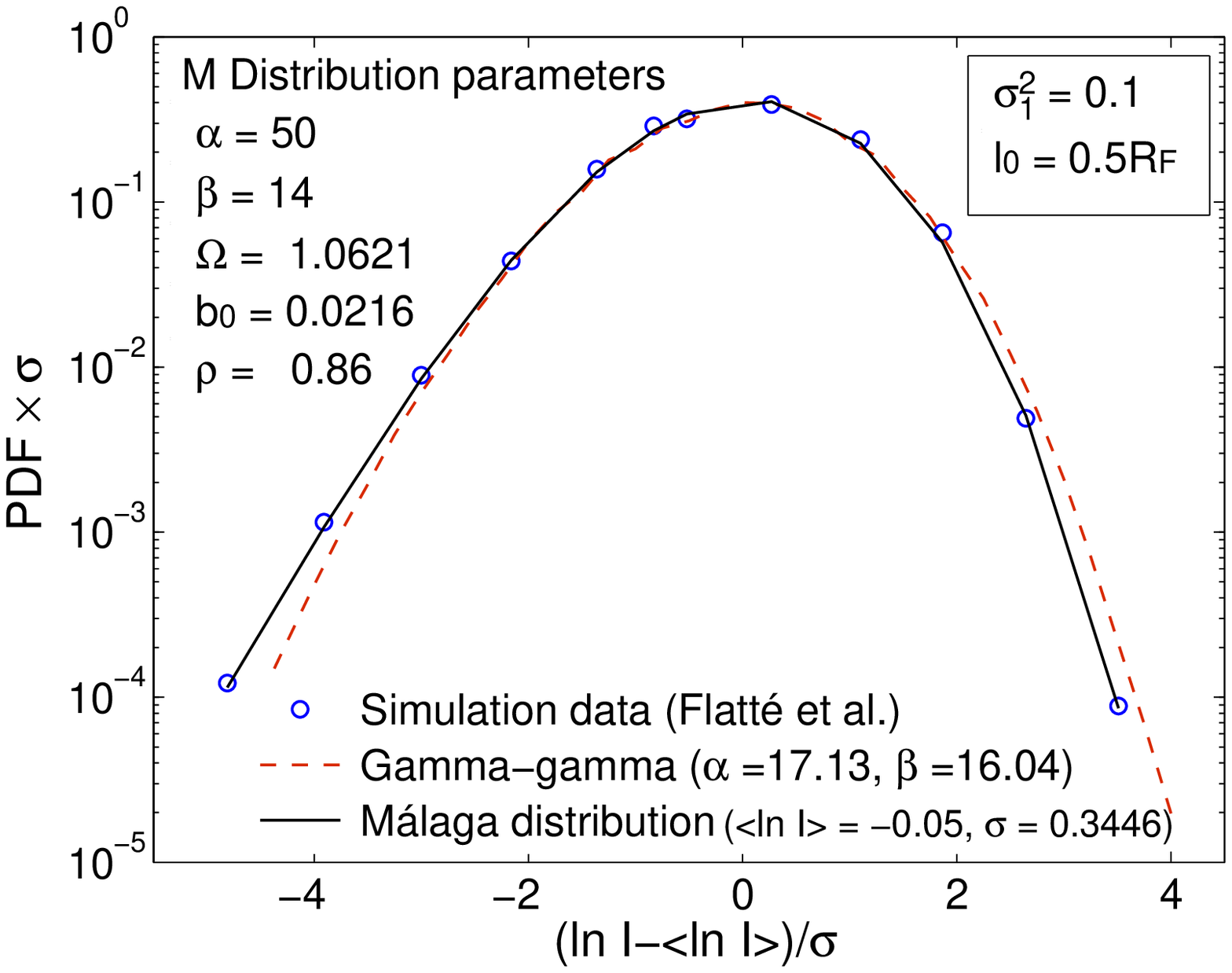}
        }
       % \hfill
        \subfigure[]{
        \includegraphics[width=4.12cm,height=3.2cm]{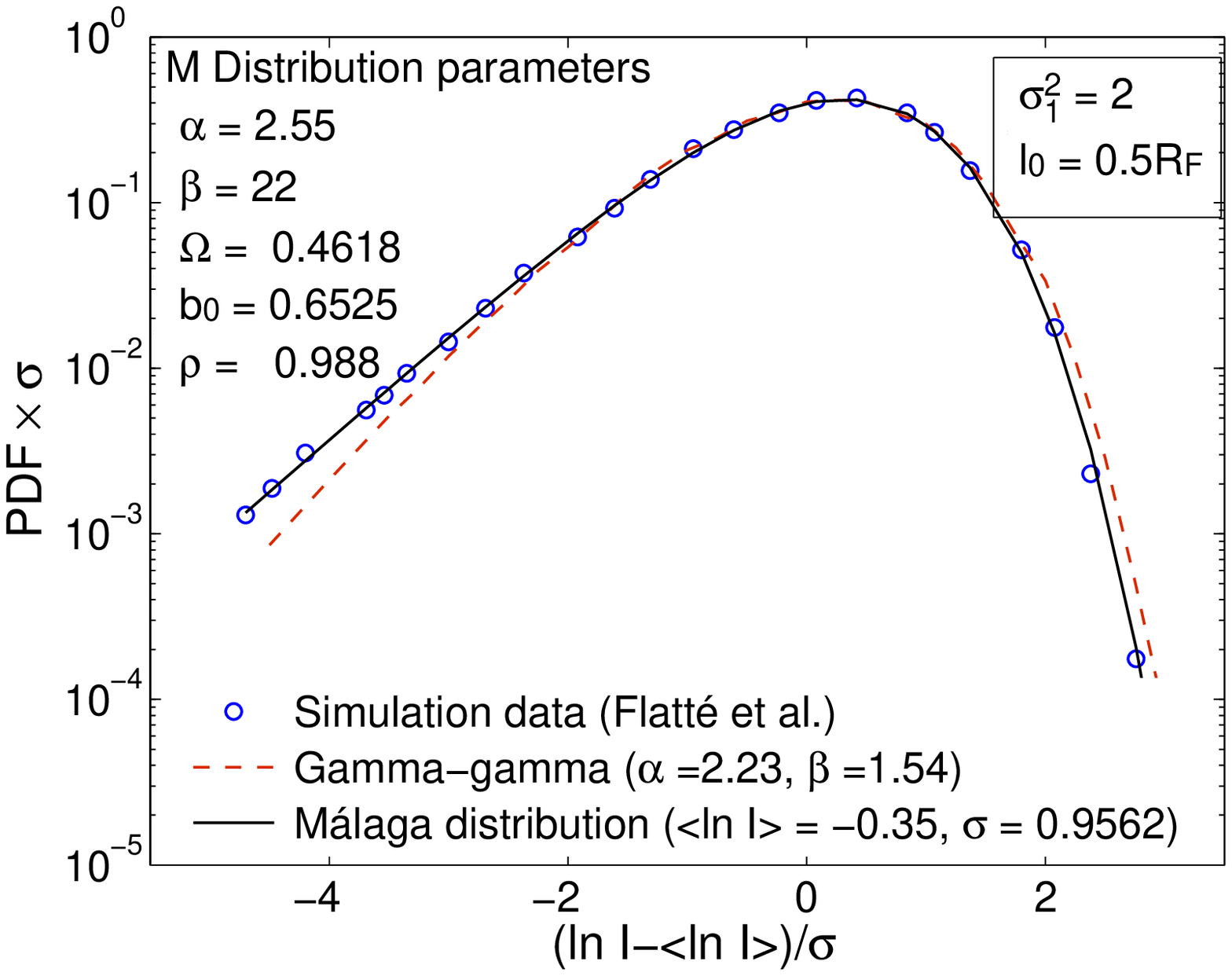}
        }
        \vfill
        \subfigure[]{
        \includegraphics[width=4.12cm,height=3.2cm]{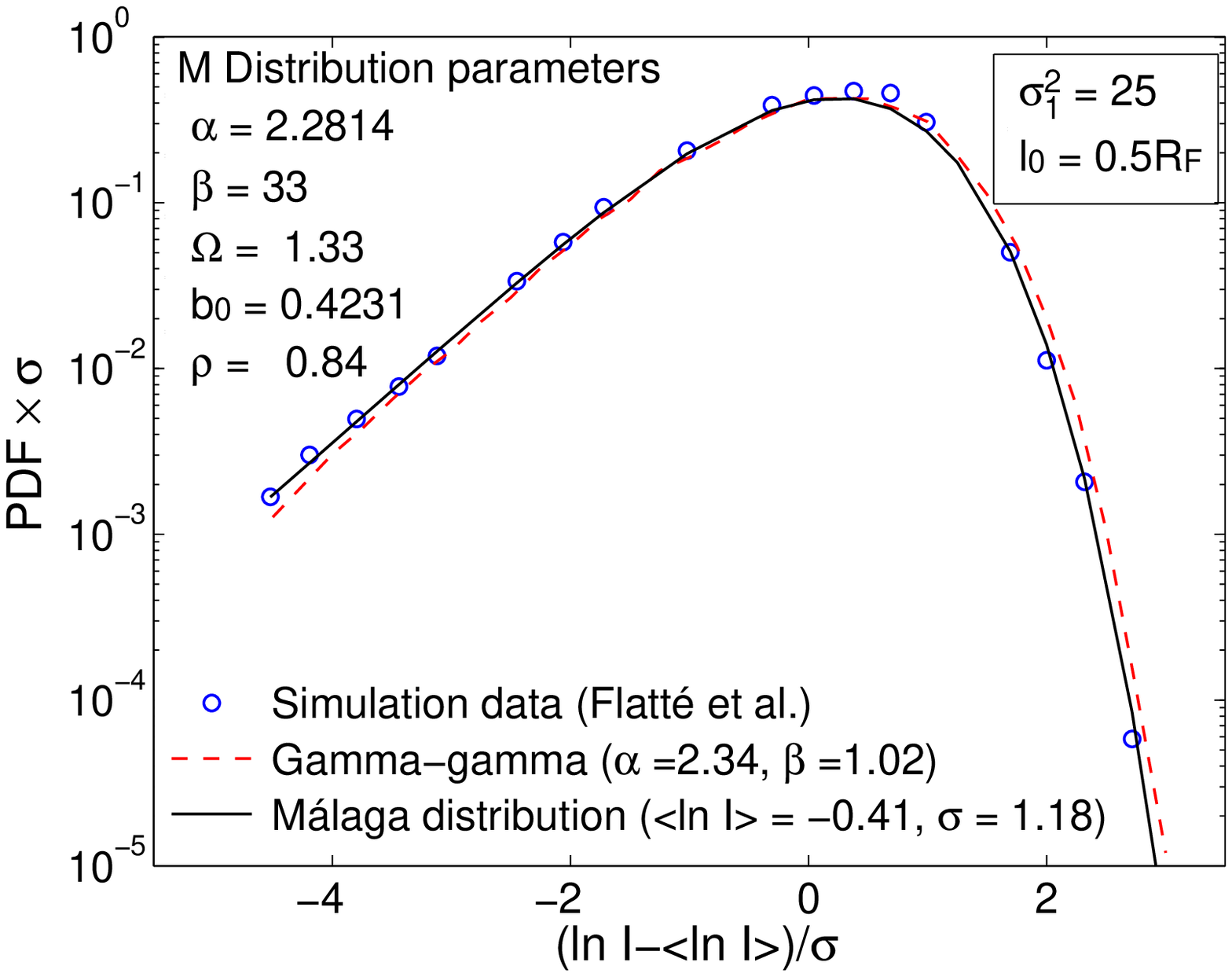}
        }
       % \hfill
        \subfigure[]{
        \includegraphics[width=4.12cm,height=3.2cm]{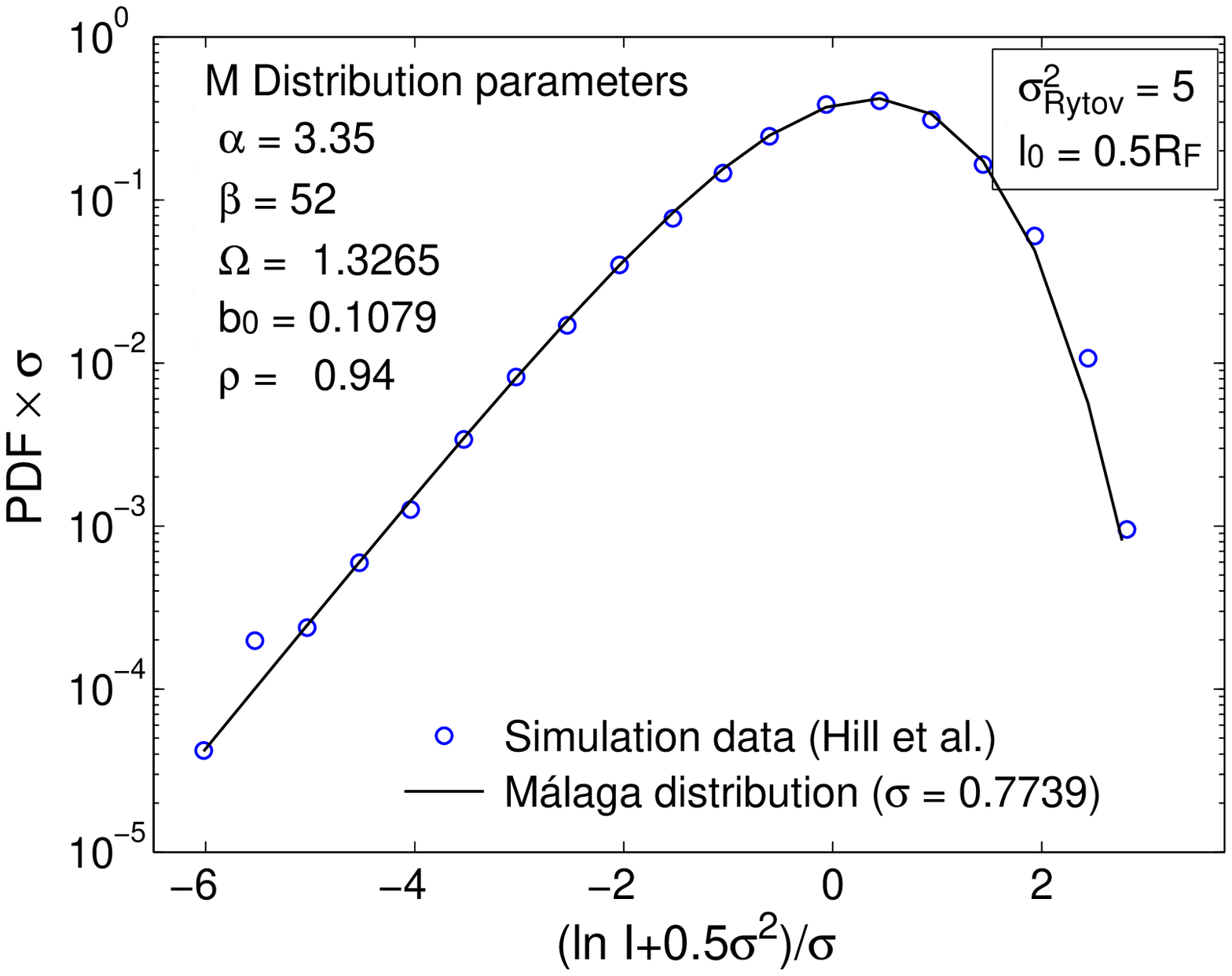}
        }
%        \caption{}
        \caption{The pdf of the scaled log-irradiance for a plane wave (Figures (a), (b) and (c)) and a spherical wave (Figure (d)) in the case of:
        (a) weak irradiance fluctuations ($\sigma_1^2=0.1$ and $l_0/R_F=0.5$); (b) moderate irradiance fluctuations ($\sigma_1^2=2$ and $l_0/R_F=0.5$);
        (c) strong irradiance fluctuations ($\sigma_1^2=25$ and $l_0/R_F=0.5$); and
        (d) strong irradiance fluctuations ($\sigma_{Rytov}^2=5$ and $l_0/R_F=0.5$).
        The blue open circles represent simulation data, the dashed red line is from the gamma-gamma pdf with $\alpha$ and $\beta$
        predicted in \cite{Fla94} and the solid black line is from our ${\rm{ }}{\cal M}$ distribution model.
        %In all subfigures, $\phi_A-\phi_B=\pi/2$.
        }
        \label{FigNumericalResults}
\end{figure}

In this section, we compare the ${\rm{ }}{\cal M}$ distribution model with some of the  published numerical simulation data
plots in \cite{Fla94} and \cite{Hil97} of the log-irradiance pdf, covering a range of conditions that extends from weak
irradiance fluctuations far into the saturation regime. It is remarkable the evident analytical tractability by directly
employing Eq. \eqref{EqPDFZEspecial}, with a finite summation of $\beta$ terms. Some results are shown in Fig.
\ref{FigNumericalResults} emphasizing its extremely high accuracy. Thus, we plot in Figs. \ref{FigNumericalResults} (a)-(c) the
predicted log-irradiance pdf associated with the ${\rm{ }}{\cal M}$ distribution (black solid line) for comparison with some of
the simulation data illustrated in Figs. 4, 5 and 7 of \cite{Fla94}. The simulation pdf values are plotted as a function of $(\ln
I - <\ln I>)/\sigma$, as in \cite{Fla94}, where $<\ln I>$ is the mean value of the log-irradiance and $\sigma=\sqrt{\sigma_{\ln
I}^2}$,
the latter being the root mean square (rms) value of $\ln I$. %The simulation pdf's were displayed in this fashion in the hope
%that it would reveal their salient features.
For sake of brevity, and as representative of typical atmospheric propagation, we
only use the inner scale value $l_0=0.5R_F$ so we can include the effect of $l_0$ in our results, where the quantity $R_F$ is the
scale size of the Fresnel zone. We also plot the gamma-gamma pdf (red dashed line) obtained in \cite{AlH01} for the sake of
comparison. In Fig. \ref{FigNumericalResults} (a) we use a Rytov variance $\sigma_1^2=0.1$ corresponding to weak irradiance
fluctuations, in Fig. \ref{FigNumericalResults} (b) we employ $\sigma_1^2=2$ corresponding to a regime of moderate irradiance
fluctuations whereas in Fig. \ref{FigNumericalResults} (c), $\sigma_1^2$ was established to $25$ for a particular case of strong
irradiance fluctuations. Values of the scaling parameter $\sigma$ required in the plots for the ${\rm{ }}{\cal M}$ pdf are
obtained from Andrews' development \cite{AlH01} in the presence of inner scale.

Finally, in Fig. \ref{FigNumericalResults} (d) we have obtained a very good fitting to the simulation data for a spherical wave
in the case of strong irradiance fluctuations. Following Hill's representation \cite{Hil97}, the simulation pdf data and pdf
values predicted by the ${\rm{ }}{\cal M}$ distribution are displayed as a function of $(\ln I+0.5\sigma^2)/\sigma$, where
$\sigma$ was defined above. In this particular case of propagating a spherical wave, an additional parameter is needed: the Rytov
parameter, $\sigma_{Rytov}^2$, defined as the weak fluctuation scintillation index in the presence of a finite inner scale.

For the particular case displayed in Fig. \ref{FigNumericalResults} (d), the gamma-gamma pdf does not fit with the simulation
data and, even more, the Beckmann pdf did not lend itself directly to numerical calculations and so are omitted. Nevertheless,
the ${\rm{ }}{\cal M}$ pdf shows very good agreement with the data once again, with the advantage of a simple functional form,
emphasized by the fact that its $\beta$ parameter is a natural number, which leads to a closed-form representation.

\section{Concluding remarks}
In this paper, a novel statistical model for atmospheric optical scintillation is presented. Unlike other models, our proposal
appears to be applicable for plane and spherical waves under all conditions of turbulence from weak to super strong in the
saturation regime. The proposed model unifies in a closed-form expression the existing models suggested in the bibliography for
atmospheric optical communications (see Table \ref{TableResumen}). In addition to the mathematical expressions and developments,
we have introduced a different perturbational propagation model
 that gives a physical sense to such existing models. Hence, the received optical intensity
is due to three different contributions: first, a line of sight component, second, a coupled-to-LOS scattering component, as a
remarkable novelty in the model, that includes the fraction of power traveling very closed to the line of sight, and eventually
suffering from almost the same random refractive index variations than the LOS component; and third, the scattering component
affected by refractive index fluctuations completely different to the other two components. The first two components are governed
by a gamma distribution whereas the scattering component is depending on a circular Gaussian complex random variable. All of them
let us model the amplitude of the irradiance (small-scale fluctuations), while the multiplicative perturbation that represents
the large-scale fluctuations is approximated to follow a gamma distribution.

%Moreover, in Table \ref{TableResumen}, we will derive some of the distribution models most frequently employed in the
%bibliography by properly choosing the magnitudes of the parameters involving the ${\rm{ }}{\cal M}$ distribution.  So, in
%conclusion, the ${\rm{ }}{\cal M}$ distribution model unifies most of the proposed statistical model for the irradiance
%fluctuations derived in the bibliography. First, ${\rm{ }}{\cal M}$ distribution makes computations extremely easy due to its
%closed-form expression. And second, parameter value $\alpha$ is directly related to calculated values of large-scale
%scintillation that depend only on values of atmospheric parameters.

Finally, we have made a number of comparisons with published plane wave and spherical wave simulation data over a wide range of
turbulence conditions (weak to strong) that includes inner scale effects. The ${\rm{ }}{\cal M}$ distribution model is found to
provide an excellent fit to the simulation data in all cases tested making computations extremely easy due to its closed-form
expression.

\section*{Appendix A: proof of Equation \eqref{EqPDFZEspecial}}
To prove Eq. \eqref{EqPDFZEspecial}, we can obtain the Laplace transform, ${\rm{ }}{\cal L}[f_Y \left( y \right);s]$, of the
shadowed Rice single pdf, $f_Y(y)$, written in Eq. \eqref{EqPDFY2}, in a direct way, with the help of Eq. (7) of Ref.
\cite{Abd03}, since the moment generating function (MGF) and the Laplace transform of the pdf $f_Y(y)$ are related by
\mbox{$M[f_Y (y);-s] = {\rm{ }}{\cal L}[ f_Y (y); s]$}:

\begin{equation}
        \begin{split}
            {\rm{ }}{\cal L}[f_Y \left( y \right);s] = \left( {\frac{{\gamma \beta }}{{\gamma \beta  + \Omega^{'}}}} \right)^\beta
\frac{{\left( {1 + \gamma s} \right)^{\beta  - 1} }}{{\left( {\frac{{\gamma \beta }}{{\gamma \beta  + \Omega^{'}}} + \gamma s}
\right)^\beta  }} = \\
=\frac{1}{\gamma }\left( {\frac{{\gamma \beta }}{{\gamma \beta  + \Omega^{'}}}} \right)^\beta \frac{{\left( {\frac{1}{\gamma } +
s} \right)^{\beta  - 1} }}{{\left( {\frac{\beta }{{\gamma \beta  + \Omega^{'}}} + s} \right)^\beta  }}.
        \end{split}
        \label{EqLaplaceTransfPDFY}
\end{equation}

Now, let us consider the following Laplace-transform pair
\begin{equation}
        \begin{split}
            \Gamma \left( {\nu {\rm{ + 1}}} \right)\left( {s - \lambda } \right)^n \left( {s - \mu } \right)^{ - \nu  -
                1}  \Leftrightarrow n!t^{\nu  - n} e^{\mu t} L_n^{\nu  - n} \left[ {\left( {\lambda  - \mu } \right)t}
                \right], \\
                {\mathop{\rm Re}\nolimits} (\nu ) > n - 1;
        \end{split}
        \label{EqLaplaceTransformPair}
\end{equation}
\noindent given in \cite{Eld54}, Eq. (4) in pp. 238, where the minor error in the sign of the argument of the Laguerre polynomial
found and corrected in \cite{Par10} has already taken into account. If we denote $\lambda = - \frac{1}{\gamma }$, $\mu  =  -
\frac{\beta }{{\gamma \beta +\Omega^{'}}}$, $n=\beta - 1$ and $\nu  = \beta - 1 $, then
\begin{equation}
        \begin{split}
            \left( {\beta  - 1} \right)!\left( {s + \frac{1}{\gamma }} \right)^{\beta  - 1}
            \left( {s + \frac{\beta }{{\gamma \beta  + \Omega^{'}}}} \right)^{ - \beta }
            \Leftrightarrow \\
            \Leftrightarrow \left( {\beta  - 1} \right)!e^{ - \frac{\beta }{{\gamma \beta  + \Omega^{'}}}t} L_{\beta  - 1}
            \left[ {\frac{{ - \Omega^{'}}}{{\gamma \beta  + \Omega^{'}}}\frac{t}{\gamma }} \right],
        \end{split}
        \label{EqLaplaceTransfPDFY2}
\end{equation}
\noindent where $L_n[\cdot]$ is the Laguerre polynomial of order $n$. If we substitute Eq. \eqref{EqLaplaceTransfPDFY2} into Eq.
\eqref{EqLaplaceTransfPDFY}, then the pdf of $Y$ can be expressed as:
\begin{equation}
        \begin{split}
         f_Y \left( y \right) = \frac{1}{\gamma }\left( {\frac{{\gamma \beta }}
         {{\gamma \beta  + \Omega^{'}}}} \right)^\beta  \exp{\left( - \frac{\beta }{{\gamma \beta  + \Omega^{'}}}y\right)}\\
         \times L_{\beta  - 1} \left[ {\frac{{ - \Omega^{'}y}}{\left({\gamma \beta  + \Omega^{'}}\right)\gamma}}
         \right].
        \end{split}
        \label{EqLaplaceTransfPDFY3}
\end{equation}

Now, to obtain the unconditional ${\rm{ }}{\cal M}$ distribution, from Eqs. \eqref{EqPDFX} and \eqref{EqLaplaceTransfPDFY3}, we
can form:
\begin{equation}
        \begin{split}
        f_I(I)&=\int_0^{\infty}{f_Y(I|x)f_X(x)\hbox{d}x} = \frac{{\alpha ^\alpha  }}
        {{\Gamma \left( \alpha  \right)}}\frac{1}{\gamma } \\
        &\times \left( {\frac{{\gamma \beta }}
        {{\gamma \beta  + \Omega^{'}}}} \right)^\beta
        \int_0^\infty  {\frac{1}{x}\exp{\left( - \frac{\beta }{{\gamma \beta  + \Omega^{'}}}\frac{I}{x}\right)}} \\
        & \times {L_{\beta  - 1} \left[ {\frac{-\Omega^{'}}{{\gamma \beta  + \Omega^{'}}}\frac{I}{{\gamma x}}} \right]
        x^{\alpha  - 1}
        \exp \left( { - \alpha x} \right)dx}.
             \label{EqPDFZ4}
        \end{split}
    \end{equation}

    By expressing the Laguerre polynomial in a series,
    \begin{equation}
        \begin{split}
         L_n \left[ x \right] = \sum\limits_{k = 0}^n {( - 1)^k
         \left(\hspace{-.15cm} {\begin{array}{*{20}c}
                                n  \\
                                k  \\
                 \end{array}}
         \hspace{-.15cm}\right)}
         \frac{{x^k }}{{k!}},
         \label{EqLaguerre}
        \end{split}
    \end{equation}
    \noindent as was shown in Eq. (8.970-1) of Ref. \cite{Gra00}, %it follows that Eq. \eqref{EqPDFZ4} becomes
%    \begin{equation}
 %       \begin{split}
  %          f_I \left( I \right) = &\frac{{\alpha ^\alpha  }}{{\Gamma \left( \alpha  \right)}}\frac{1}{\gamma }\left(
   %         {\frac{{\gamma \beta }}{{\gamma \beta  + \Omega^{'}}}} \right)^\beta \\
    %        &\times \sum\limits_{k = 1}^\beta  {( - 1)^{k - 1}
     %       \left(\hspace{-.15cm} {\begin{array}{*{20}c}
      %                     {\beta  - 1}  \\
       %                    {k - 1}  \\
        %            \end{array}}
         %   \hspace{-.15cm}\right)}
          %  \frac{1}{{(k - 1)!}}\left( {\frac{{ - \Omega^{'}}}{{\gamma \beta  + \Omega^{'}}}\frac{1}{\gamma }} \right)^{k - 1}\\
           % &\times I^{k - 1} \cdot\int_0^\infty  {\exp{\left( - \frac{\beta }{{\gamma \beta  + \Omega^{'}}}\frac{I}{x}- \alpha x\right)} x^{\alpha  - 1 - k}
            %dx}.
             %\label{EqPDFZ5}
%        \end{split}
 %   \end{equation}
%
 %   Now, we denote by $G_k$ the integral:
  %  \begin{equation}
   %     \begin{split}
    %          G_k  = \int_0^\infty  {x^{\alpha  - 1 - k} \exp \left( { - \frac{\beta }{{\gamma \beta  + \Omega^{'}}}\frac{I}{x} - \alpha x}
     %         \right)dx}.
      %       \label{EqGk}
       % \end{split}
%    \end{equation}
%
%    Now,
    and using Eq. (3.471-9) of Ref. \cite{Gra00},
%    \begin{equation}
%        \begin{split}
%            \int_0^\infty  {x^{\nu  - 1} \exp{\left( - \frac{\beta }{x} - \gamma x\right)} dx = 2\left( {\frac{\beta }{\gamma }}
%            \right)^{\frac{\nu }{2}} } K_\nu  \left( {2\sqrt {\beta \gamma } } \right),
%             \label{EqIntegralGrah}
%        \end{split}
%    \end{equation}
% \noindent   we can solve $G_k$:
it follows that Eq. \eqref{EqPDFZ4} becomes

%    \begin{equation}
%        \begin{split}
%             G_k  = %2\left( {\frac{{\frac{{\beta z}}{{\gamma \beta  + \Omega^{'}}}}}{\alpha }} \right)^{\frac{{\alpha  - k}}{2}} K_{\alpha  - k}
%%             \left( {2\sqrt {\frac{{\beta z}}{{\gamma \beta  + \Omega^{'}}}\alpha } } \right) =
%             2\left( {\frac{\beta }{{\alpha \left( {\gamma \beta  + \Omega^{'}} \right)}}} \right)^{\frac{{\alpha  - k}}{2}}
%             I^{\frac{{\alpha  - k}}{2}} K_{\alpha  - k} \left( {2\sqrt {\frac{{\beta I}}{{\gamma \beta  + \Omega^{'}}}\alpha } }
%             \right).
%             \label{EqGk2}
%        \end{split}
%    \end{equation}
%
%    Employing this latter result and inserting it into Eq. \eqref{EqPDFZ5}, we find the pdf of $I$ in the form:
    \begin{equation}
        \begin{split}
            f_I \left( I \right) = &\frac{{2\alpha ^{\frac{\alpha }{2}} }}
            {{\Gamma \left( \alpha  \right)}}\frac{1}{{\gamma ^{1 + \frac{\alpha }{2}} }}
            \left( {\frac{{\gamma \beta }}{{\gamma \beta  + \Omega^{'}}}} \right)^\beta
            \left( {\frac{{\gamma \beta  }}{{\gamma \beta  + \Omega^{'}}}} \right)^{\frac{\alpha }{2}}  \\
            &\times \sum\limits_{k = 1}^\beta  {
            \left(\hspace{-.15cm} {\begin{array}{*{20}c}
                       {\beta  - 1}  \\
                       {k - 1}  \\
                    \end{array}}
            \hspace{-.15cm}\right)}
            \frac{1}{{k - 1!}}\left( {\frac{{ \Omega^{'}}}{{\gamma\left(\gamma \beta  + \Omega^{'}\right)}}} \right)^{k - 1}\\
            & \times \left( {\frac{\beta }{{\alpha \left( {\gamma \beta  + \Omega^{'}} \right)}}} \right)^{ - \frac{k}{2}}
             I^{\frac{{\alpha  + k}}{2} - 1} K_{\alpha  - k} \left( {2\sqrt {\frac{{\alpha \beta I}}{{\gamma \beta  + \Omega^{'}}}} }
            \right).
             \label{EqPDFZ6}
        \end{split}
    \end{equation}

    Finally, Eq. \eqref{EqPDFZ6} can be rewritten as Eq. \eqref{EqPDFZEspecial}, as was already anticipated in Section III.

\section*{Appendix B: proof of Equation \eqref{EqMomentsZEspecial}}
From Eq. \eqref{EqComponentesIrradiancia}, the observed irradiance, $I$, of our proposed propagation model can be expressed as:
$I=XY$, where the pdf of variables $X$ and $Y$ were written in Eqs. \eqref{EqPDFX} and \eqref{EqPDFY2}, respectively. Based on
assumptions of statistical independence for the \mbox{underlying random processes, $X$ and $Y$, then:}
\begin{equation}
        \begin{split}
                 m_k \left( I \right) = E\left[ {X^k } \right]E\left[ {Y^k } \right] = m_k \left( X \right)m_k \left( Y
                 \right).
             \label{EqMomentsZ2}
        \end{split}
    \end{equation}

From Eq. (2.23) of Ref. \cite{Sim05}, the moment of a Nakagami-m pdf is given by:
\begin{equation}
        \begin{split}
                 m_k \left( X \right) = \frac{{\Gamma \left( {\alpha  + k} \right)}}{{\Gamma \left( \alpha  \right)\alpha ^k
                 }}.
             \label{EqMomentsX}
        \end{split}
    \end{equation}
\noindent On the other hand, from Eqs. \eqref{EqLaplaceTransfPDFY3} and \eqref{EqLaguerre}, %we can obtain the moment of the
%Rician-shadowed distribution, given by:
%\begin{equation}
%        \begin{split}
%                  m_k \left( Y \right) = &
%                   \frac{1}{\gamma }\left( {\frac{{\gamma \beta }}{{\gamma \beta  + \Omega^{'}}}} \right)^\beta
%                  \sum\limits_{r = 0}^{\beta  - 1}
%                  {\beta-1 \choose r} \hspace{-.1cm}
%                  \left( {\frac{\Omega^{'}}{{\gamma \left( {\gamma \beta  + \Omega^{'}} \right)}}} \right)^r \\
%                  & \times \frac{1}{{r!}}
%                  \int_0^\infty  {y^{k + r} \exp{\left( - \frac{\beta y}{{\gamma \beta  + \Omega^{'}}}\right)} } dy.
%             \label{EqMomentsYEspecial}
%        \end{split}
%    \end{equation}
%
%Now, from
\noindent and employing Eq. (3.381-4) of Ref. \cite{Gra00},
%\begin{equation}
%        \begin{split}
%            \int_0^\infty  {x^{\nu  - 1} \exp{\left( - \mu x \right)} dx} = {1\over \mu^{\nu}}\Gamma\left(\nu\right),
%            \hspace{0.75cm} \\
%            [\hbox{Re}(\mu)>0,  \hbox{Re}(\nu)>0],
%             \label{EqIntegralGrah2}
%        \end{split}
%    \end{equation}
%\noindent we can express Eq. \eqref{EqMomentsYEspecial} as:
\noindent obtain the moment of the Rician-shadowed distribution, given by:
\begin{equation}
        \begin{split}
                  m_k \left( Y \right) = &\frac{1}{\gamma }\left( {\frac{{\gamma \beta }}{{\gamma \beta  + \Omega^{'}}}} \right)^\beta
                  \sum\limits_{r = 0}^{\beta  - 1} {
                  \left(\hspace{-.15cm} {\begin{array}{*{20}c}
                           {\beta  - 1}  \\
                           r  \\
                           \end{array}}
                  \hspace{-.15cm}\right)}
                  \frac{1}{{r!}} \\
                  & \times\left( {\frac{\Omega^{'}}{{\gamma \left( {\gamma \beta  + \Omega^{'}} \right)}}} \right)^r
                 \frac{{\Gamma \left( {k + r + 1} \right)}}{{\left( {\frac{\beta }{{\gamma \beta  + \Omega^{'}}}} \right)^{k + r +
                1}}}.
             \label{EqMomentsYEspecial2}
        \end{split}
    \end{equation}
Finally, when performing the product of Eq. \eqref{EqMomentsX} by Eq. \eqref{EqMomentsYEspecial2}, we certainly obtain Eq.
\eqref{EqMomentsZEspecial}.

%\vspace{.3cm} \noindent\emph{Acknowledgement}: This work was supported by the Spanish Ministerio de Educaci\'on y
%Ciencia, Project TEC2008-06598.

%\pagebreak
%\IEEEtriggeratref{11}
%Bibliografía
\def\baselinestretch{1}

% FIGURAS EN FLOTANTE
%\clearpage
%
%\section*{List of Figure Captions}
%
%\noindent Fig. 1. Scintillation sequences generated from (a) an AR model; (b) a space-time separable statistics
%model, for a system with three receivers. The EGC sequence is displayed in thicker solid line. The wind velocity
%is established to $u_{\perp}=20 $ m/s.
%
%\bigskip
%
%\noindent Fig. 2. Burst error rate of a system with 3 receivers versus normalized average optical power using OOK
%and ML detection, for different values of $\rho_{ij}$ and $\sigma_{\chi}^2$, with a wind velocity of $u_{\perp}=20
%$ m/s. The burst error length is established to (a) 192 bits, (b) 64 bits.
%
%\bigskip
%
%\noindent Fig. 3. Burst error rate of a system with 3 receivers versus normalized average optical power using OOK
%format, EGC and ML detection, assuming $\rho_{12}=0.56$ and $\rho_{13}=0.1$, being $u_{\perp}=20 $ m/s, whereas
%the burst error length is established to 192 bits. A OOK-GS format with d.c. of $100\%$ is represented in (a), for
%different values of $\sigma_{\chi}^2$. Different transmission formats have been displayed in (b) for
%$\sigma_{\chi}^2=0.15$.
%
%%\listoffigures
%\clearpage

%\pagebreak
\end{document}